\documentclass[twocolumn, showpacs,preprintnumbers,amsmath,amssymb]{revtex4}
\usepackage{amssymb}
\usepackage{xcolor}
\usepackage{graphicx}
\usepackage{dcolumn}
\usepackage{bm}
\usepackage{multirow}
\usepackage{booktabs}
\usepackage{natbib}
\usepackage{soul}
\usepackage{amssymb}
\usepackage{amsmath}	
\usepackage{color}
\usepackage{cases} 

\newcommand{\Vmat}{{\boldmath $\cal V$}}
\newcommand{\Kmat}{{\boldmath $\cal K$}}

\newcommand{\numat}{\mbox{\boldmath $\nu$}}

\newcommand{\Smat}{\mbox{\boldmath $\cal S$}}

\newcommand{\Xmat}{\mbox{\boldmath $X$}}

\begin{document}
\preprint{APS/123-QED}

\title{Dissociative recombination of NS$^{+}$ in collisions with slow electrons}

\author{R. Hassaine$^{1}$}\email[]{riyad.hassaine@etu.univ-lehavre.fr}
\author{F. Gauchet$^{1}$}
\author{F. Iacob$^{2}$}
\author{J. Zs Mezei$^{1,3}$}\email[]{mezei.zsolt@atomki.hu}
\author{E. Roueff$^{4}$}
\author{J. Tennyson$^{1,3,5}$}
\author{I. F. Schneider$^{1,6}$}\email[]{ioan.schneider@univ-lehavre.fr}
\affiliation{$^{1}$LOMC-UMR6294, CNRS, Universit\'e Le Havre Normandie, 76600 Le Havre, France}%
\affiliation{$^{2}$Physics Faculty, West University of Timisoara, 300223,Timisoara, Romania}%
\affiliation{$^{3}$HUN-REN Institute for Nuclear Research (ATOMKI), H-4001 Debrecen, Hungary}%
\affiliation{$^{4}$LERMA CNRS-UMR8112, Observatoire de Paris, Universit\'e PSL, F-92190, Meudon, France}%
\affiliation{$^{5}$Department of Physics and Astronomy, University College London, WC1E 6BT London, UK}%
\affiliation{$^{6}$LAC-UMR9188, CNRS Universit\'e Paris-Saclay, F-91405 Orsay, France}%
\date{\today}

\begin{abstract}
Cross sections and rate coefficients for the Dissociative Recombination (DR) of the NS$^{+}$ ion induced by collisions with low-energy electrons are reported for temperatures between 10 and 1000 K, relevant to a large range of  interstellar cloud temperatures. Uncertainties are discussed for these rates. Comparisons are made with DR rates for the isovalent NO$^+$ molecular ion which are found to be much faster.  The present findings lead to a moderate dissociative reaction rate coefficient, smaller by a factor of 2 than the current estimates reported in the different kinetic databases for a temperature of 10 K. We consider that our rate coefficients obtained through multichannel quantum defect theory for NS$^{+}$ are likely to be better than those displayed in the different kinetic databases.
\end{abstract}


\maketitle

\section{Introduction}

Nitrogen sulfide (NS) was first detected in the early years of radio astronomy toward `SgrB2 interstellar cloud~\cite{Gottlieb1975} and its presence has later been found in giant molecular clouds~\cite{McGonagle1997}, dense cores~\cite{HilyBlant2022} as well as in comets~\cite{Irvine2000}. 
Recently, new observations have led to the discovery of the  nitrogen sulfide cation NS$^{+}$ thanks to laboratory spectroscopy~\cite{Cernicharo2018}; the ion is found to be ubiquitous, from cold molecular cores, prestellar cores, and to shocks \cite{Cernicharo2018} and photon dominated regions~\cite{Riviere2019}.
Dissociative recombination (DR): 
\begin{equation}
\label{eq:DR} 
\mbox{NS}^{+}(v_{i}^{+})+e^{-}(\varepsilon) 
\longrightarrow 
\mbox{N} + \mbox{S}
\end{equation}
\noindent
- where $v_{i}^{+}$ is the initial vibrational number of the target and $\varepsilon$
is the energy of the incident electron -
is likely
the major removal mechanism for the ion. 
It competes with autoinization, resulting in vibrational transitions: 
\begin{equation}
\label{eq:SEC} 
\mbox{NS}^{+}(v_{i}^{+})+e^{-}(\varepsilon) 
\longrightarrow 
\mbox{NS}^{+}(v_{f}^{+})+e^{-}({\varepsilon}')
\end{equation}
\noindent  
i.e. vibrational excitation - when $v_{f}^{+}$ is larger than $v_{i}^{+}$
and $\varepsilon$ exceeds the threshold of reaching this excited level - or 
vibrational de-excitation - when $v_{f}^{+}$ is smaller than $v_{i}^{+}$ - $\varepsilon'$
being the energy of the scattered electron.

Here computations are presented which were performed using our step-wise Multichannel Quantum Defect Theory (MQDT)~\cite{Giusti1980,mezei2019} and references therein.  We focus on calculating dissociative recombination (eq.~(\ref{eq:DR})) cross sections and rate coefficients for collision energies/temperatures up to 1 eV/1000 K. The first attempt to provide uncertainties for our MQDT method is presented. Due to the lack of other theoretical or experimental results we will compare our results with those obtained for NO$^+$, another closed shell molecular cation relevant for the upper atmosphere of Earth, that has similar electron structure and chemical properties to NS$^+$. 

The organisation of the paper is as follows: In section~\ref{sec:theory}, a succinct presentation of our theoretical approach is given. Section~\ref{sec:results} contains the major results - cross sections and rate coefficients - and section~\ref{sec:conclusions} summarizes the concluding remarks.

\section{Theoretical Method}{\label{sec:theory}}

The effectiveness of our theoretical approach based on the step-wise MQDT~\cite{Giusti1980,mezei2019} for modeling electron/diatomic cation collisions has been demonstrated through numerous previous studies involving various species. Examples of these include H$_2^+$ and its isotopologues, where we account for rotational and isotope effects across low to medium-high collision energies \cite{motapon2008,waffeutamo2011,chakrabarti2013,motapon2014,Epee2016,djuissi2020}. Additionally, we have explored core-excited effects for CH$^+$ \cite{mezei2019} and for SH$^+$ \cite{Kashinski2017,boffelli2023}. As the main elements of our approach have been extensively presented in previous works, such as~\cite{mezei2019},  we will focus here on only outlining its main steps.

Reactions~(\ref{eq:DR}) and (\ref{eq:SEC}) encompass \textit{ionization} channels, characterizing the interaction of an electron with the target cation, as well as \textit{dissociation} channels, relating to atom-atom scattering. The interaction of these channels gives rise to quantum interference between the \textit{direct} mechanism relying on open channels—where capture occurs into a doubly excited dissociative state of the neutral system— and the \textit{indirect} one—where capture takes place through a Rydberg bound state of the molecule pertaining to \textit{closed} channels. These Rydberg states are predissociated by the dissociative channels. In both mechanisms, autoionization, which relies on the presence of \textit{open} ionization channels, competes with predissociation, resulting in the cation's vibrational excitation or de-excitation-reaction (\ref{eq:SEC}). Each individual ionization channel is constructed by adding an electron to the NS$^+$ ion within its fundamental electronic state $X$ ${^1}\Sigma{^+}$ at a specific ($v_i^+$) vibrational level. These ionization channels interact with all available dissociation exit channels, via Rydberg-valence interaction, for all relevant molecular symmetries - $^2\Sigma^{+}$, $^2\Pi$, and $^2\Delta$, labeled by "\textit{sym}" within this article.

The MQDT approach for studying dissociative recombination relies on prior knowledge of the potential energy curves (PECs) corresponding to the ion's ground state as well as that of the dissociative states of the neutral molecule and of its mono-excited  Rydberg states. These latter states, which we can organize in Rydberg series, each of them characterized by geometry-dependent  quantum defects. Additionally, the electronic couplings between the dissociation and ionization continua (referred as Rydberg-valence couplings) are needed.

\begin{figure}
\centering
		\includegraphics[width=0.95\columnwidth]{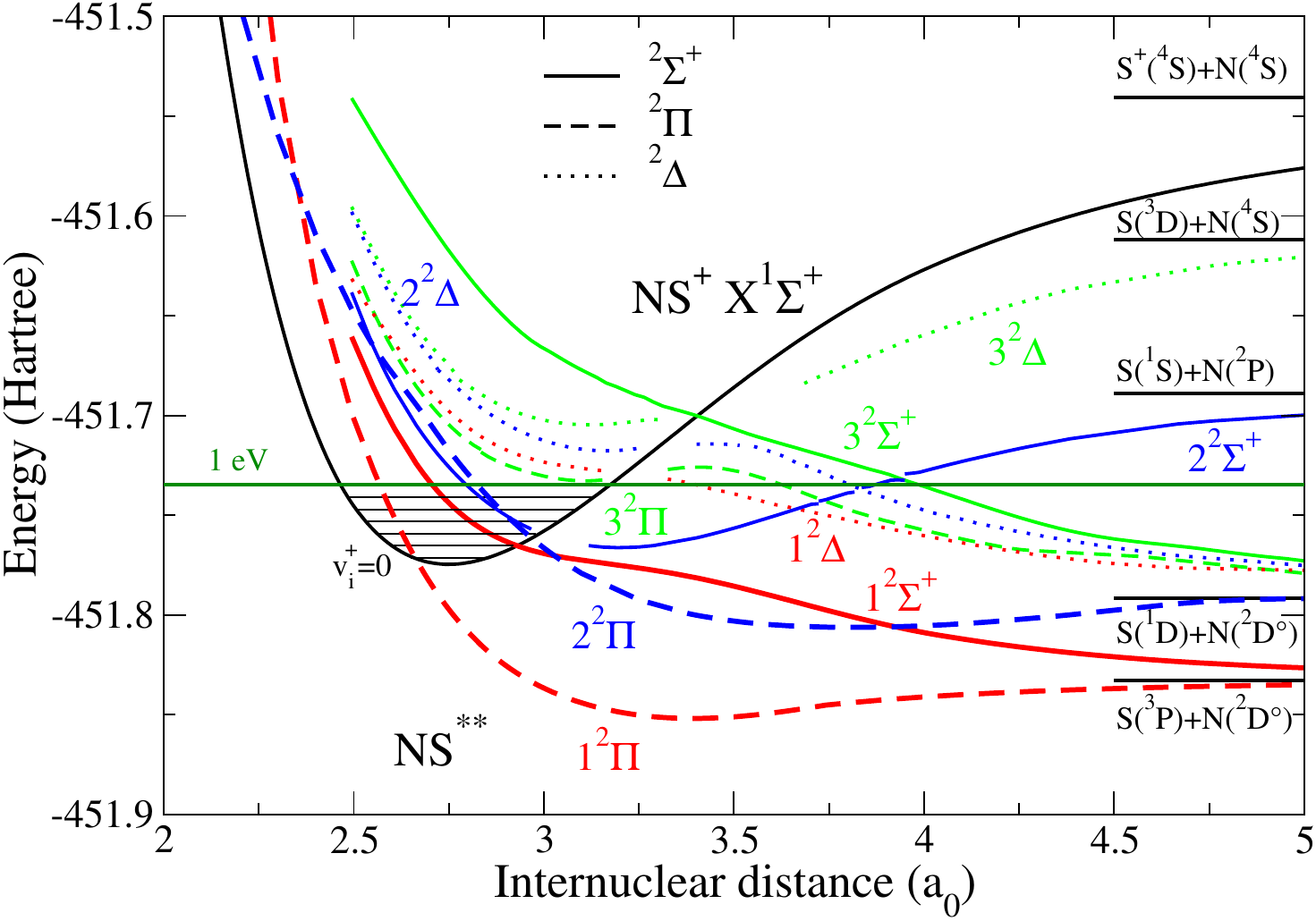}
    \caption{Energy diagram of the NS$^+$/NS molecular system, relevant for dissociative recombination, compiled from~\cite{iacob2022}. Given are the potential energy curves (PECs) and the atomic states of their corresponding asymptotic limits are given as short horizontal lines at large internuclear separation. The thick black solid curve represents the PEC of the ground electronic state of the NS$^+$ ion, while the solid colored curves represent NS$^{**}$ dissociative states computed by the R-matrix method. The thin black horizontal lines denote the lowest vibrational levels of the ion.
    }
    \label{fig:1}
\end{figure}

A summary of the relevant PECs, taken from Iacob {\it et al.} \cite{iacob2022}, can be seen in Fig.~\ref{fig:1}. The black curve stands for the ground electronic state of the cation, while the colored solid, dashed and dotted lines represent the dissociative states of NS belonging to the $^2\Sigma^+$, $^2\Pi$ and $^2\Delta$ symmetries.

The current MQDT calculations neglect rotational effects. Within each of the symmetries involved, focus is directed towards the few lowest dissociative states accessible at low energy of the incident electron - below 1 eV - relevant for the interstellar media. 
For a given total symmetry, here we consider $^2\Sigma^+$, $^2\Pi$ and $^2\Delta$,  the initial step of our methodology involves constructing the {\it interaction matrix} {\Vmat}, which serves as the driving force during the collision process. Its elements quantify the interconnections between the ionization and the dissociation channels.

The next step consists of construction of the short-range reaction matrix {\Kmat}.
This is achieved using  a second-order perturbative solution of the Lippmann-Schwinger equation. The {\Kmat}-matrix is then diagonalized, its eigenvalues being related to the long-range phase-shifts of the eigenfunctions. Applying a Cayley transformation on these matrices \cite{cayley1894collected} allows the generalized scattering matrix 
{\Xmat} to be constucted. Elimination of the closed channels~\cite{Seaton1983} is then performed, yielding the physical scattering matrix
{\Smat}:
\begin{equation}
\Smat = \Xmat_{oo}-\Xmat_{oc}\frac{1}{\Xmat_{cc}-\exp({\rm -i 2 \pi} \numat)} \Xmat_{co}\,,
\label{eq:elimination}
\end{equation}
which relies on the block-matrices built not only for the open channels, {\Xmat$_{oo}$},  but also for the 
closed ones, {\Xmat$_{oc}$, \Xmat$_{co}$} and \Xmat$_{cc}$. The diagonal matrix {\numat} appearing in the denominator of equation~(\ref{eq:elimination}) incorporates the effective quantum numbers corresponding to the vibrational thresholds of the closed ionization channels at the given total energy of the system.

\begin{figure}
\centering
\includegraphics[width=0.95\columnwidth]{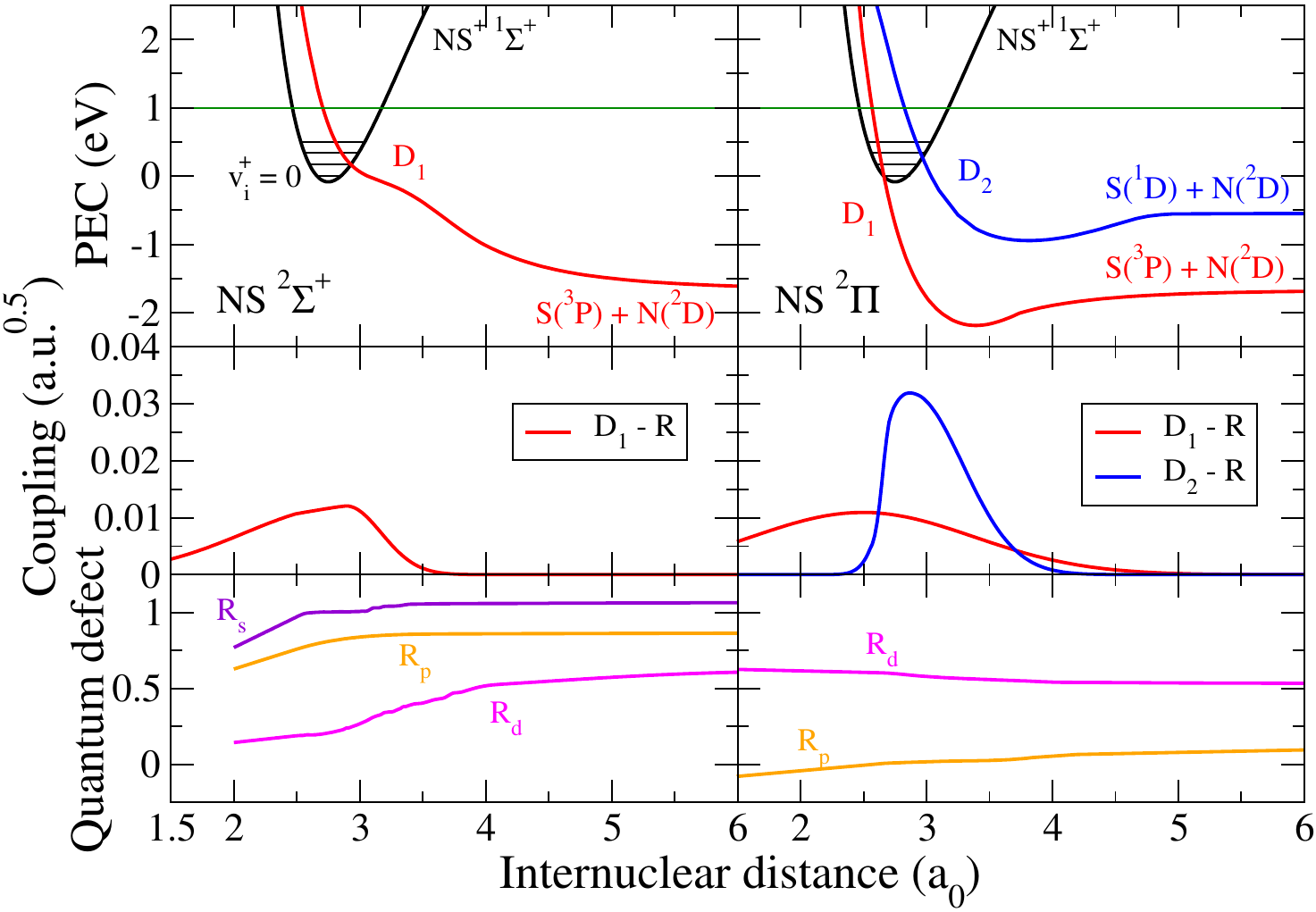}    
\caption
{Relevant molecular data sets used in the nuclear dynamics calculations. In the top row, the thick red solid curves ($D_{1}$) stand for the PEC of NS$^{**}$ $1\,^2\Sigma^{+}$ (left panel) and $1\,^2\Pi$ (right panel), while the thick blue solid curve ($D_{2})$ represents NS$^{**}$ $2\,^2\Pi$ state. The middle row displays the electronic couplings between these dissociative states and the ionization continuum. The quantum defects $\mu$ characterizing the Rydberg series of dominant partial waves $s$, $p$, and $d$ are shown in the bottom row.}
    \label{fig:2}
\end{figure}

\begin{table}\centering
	\caption{Asymptotic states of NS$^{**}$ relevant for low-energy impact collisions. The corresponding energies are given with respect to NS$^{**}$ lowest limit: S($^{3}$P)+N($^{4}$S).} 
	  \label{tab:asym}
	  \footnotesize
	\scalebox{1.0}{
	\begin{tabular}{cccccccc}
		\hline
  \rule{0pt}{3ex} 
	State & Energy (eV) & Symmetry\\
	    \hline
     \rule{0pt}{3ex}
      S($^{3}$P)+N($^{2}$D$^{o}$) & 2.381 & 1$^{2}\Sigma^{+}$, 1$^{2}\Pi$\\
      S($^{1}$D)+N($^{2}$D$^{o}$) & 3.524 & 1$^{2}\Pi$\\
      \hline
	\end{tabular}}
\end{table}
\normalsize

Finally, the global cross section for the dissociative recombination is:
 \begin{equation}
\begin{split}
\sigma _{diss \leftarrow v_{i}^{+}}  &=\frac{\pi}{4\varepsilon} \sum_{sym}\rho^{(sym)} \sum_{l,j}\mid S^{(sym)}_{d_{j},v_{i}^{+}l}\mid^2,\label{eqDR} 
\end{split}
\end{equation}		
\noindent 
where 
$\rho$ represents the ratio of multiplicities between the neutral system and the ion for a specific neutral symmetry. 
The thermal rate coefficients are determined by convoluting the cross section with the Maxwell energy distribution function of the free electrons:
\begin{equation}\label{rate}
\alpha(T) = \frac{2}{kT} \sqrt{\frac{2}{\pi mkT}} \int_{0}^{+\infty} \sigma(\varepsilon)\varepsilon \exp(-\varepsilon/kT) d\varepsilon,
\end{equation}
\noindent where $m$ is the electron's mass, $T$ the temperature, and $k$ the Boltzmann constant.
Applying the step-wise MQDT method outlined in the previous section, we have calculated the dissociative recombination (eq.~(\ref{eqDR})) cross sections of NS$^{+}$ for the lowest two vibrational levels of the ground electronic state of the cation. Since NS$^{+}$ molecular cation has been observed in the diffuse interstellar medium (ISM) for temperatures between $10$ and $300$ K~\cite{Cernicharo2018}, corresponding to electron energies in the range between $0.01$ meV and  $1$ eV, here we focus on producing cross sections for the lowest relevant open dissociative states of the neutral. 
For the ionization channels, the partial waves we took into account for the incident electron are $s$, $p$, and $d$ for the $^2\Sigma^+$ states and $p$ and $d$ for the $^2\Pi$ ones~\cite{iacob2022}.

\begin{figure}\centering
		\includegraphics[width=0.95\columnwidth]{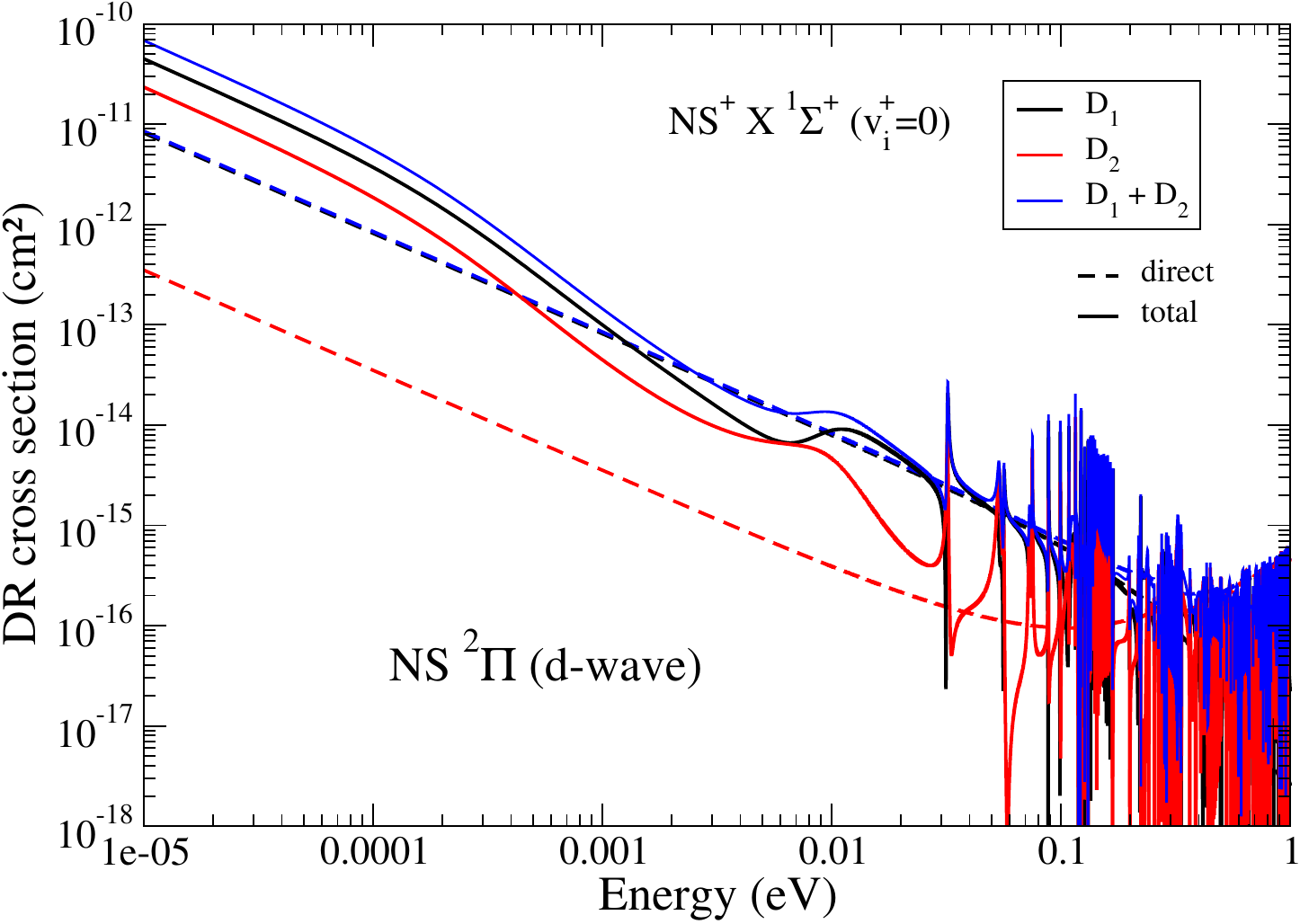}
    \caption{Dissociative recombination of NS$^+$ in its ground state with an electron into  NS states  - $D_1$ or/and $D_2$ of $^2\Pi$ symmetry - restricted to the contribution of the 
    $d$
    partial wave: Total vs direct mechanism.
    }
    \label{fig:3}
\end{figure}
\begin{figure}\centering
		\includegraphics[width=0.95\columnwidth]{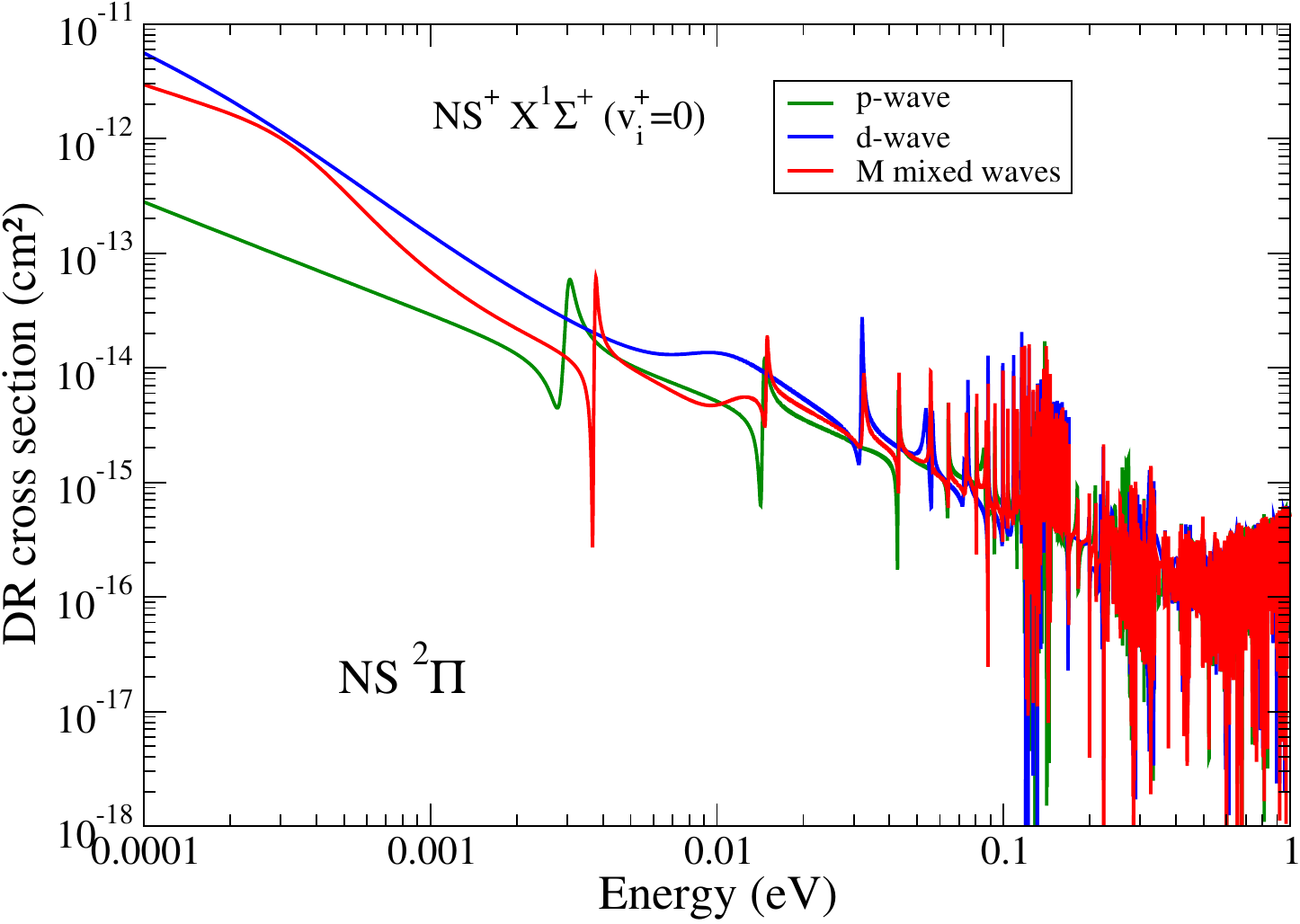}
    \caption{Dissociative recombination cross section of NS$^+$ for the $^2\Pi$ molecular symmetry for $p$, $d$ and mixed "M"$=(1/2)\times p + (1/2)\times d$ partial waves for the incident electron. 
    }
    \label{fig:4}
\end{figure}

Starting from Fig. \ref{fig:1} containing the PECs of NS$^{+}$ and doubly excited dissociative states of NS, and considering the temperature ranges relevant for the ISM we reduced the input molecular data to those curves presented in Fig.~\ref{fig:2}. We also limited our calculations to collision energies up to 1 eV shown with dark green line in Figs.~\ref{fig:1} and \ref{fig:2}. This restricts the open dissociation channels to those whose asymptotic limits are given in the first column of table~\ref{tab:asym}. In addition, the unfavourable crossing points - far from the Franck-Condon region - between the PEC's of the remaining open dissociative channels 1\ $^2\Delta$, 2\ $^2\Delta$, 3\ $^2\Sigma$, 3\ $^2\Pi$ and that of the target ion - make them irrelevant for dissociative recombination at low energy.
Consequently, $1\,^{2}\Sigma^{+}$, $1\,^{2}\Pi$, $2\,^{2}\Pi$ are the only relevant dissociative paths to be considered in this study. Figure~\ref{fig:2} contains the molecular data characterizing  these dissociative states.
The middle panels represent their electronic couplings with the ionization continuum. The quantum defects generating the complete Rydberg series corresponding to the partial waves $s$, $p$, $d$ for $^{2}\Sigma^{+}$ symmetry and $p$, $d$ for $^{2}\Pi$ symmetry are displayed in the bottom panels.

 \begin{figure*}\centering
		\includegraphics[width=0.95\columnwidth]{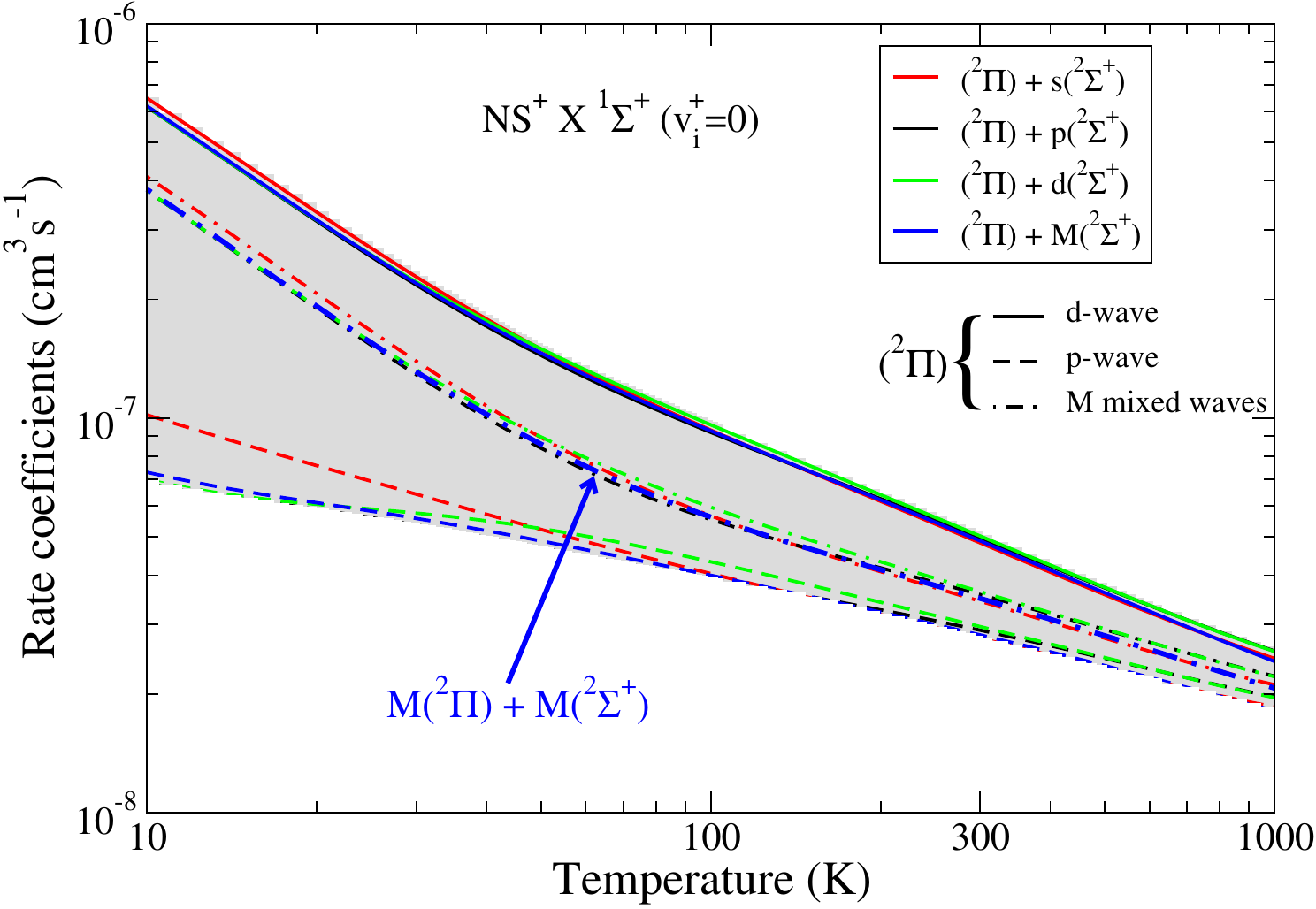}
		\includegraphics[width=0.95\columnwidth]{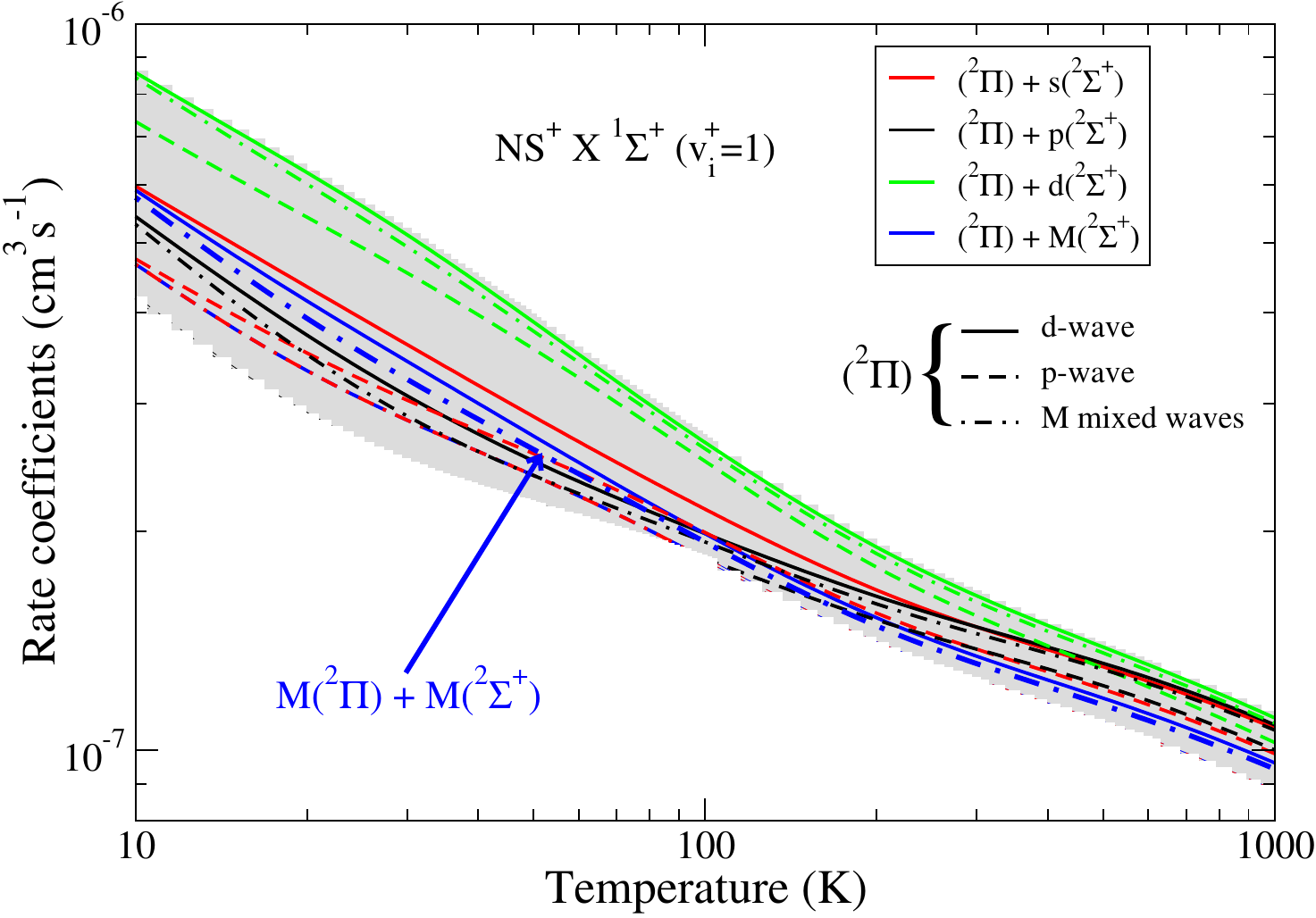}
    \caption{Rate coefficients for dissociative recombination of NS$^+$ cation from its lowest two vibrational levels and  uncertainties due to the use of different partial waves for the incident electrons. Curves give the partial wave contributions for the incident electron. "M" denotes mixed partial waves (($1/2)\times p + (1/2)\times d$ for $^{2}\Pi$ states and $(1/3)\times s + (1/3)\times p + (1/3)\times d$ for $^{2}\Sigma^{+}$ states). The lower limit of the gray band stands for the minimum rate coefficient as a function of temperature while its upper limit stands for the maximum rate coefficient as a function of temperature. 
    }
    \label{fig:5}
\end{figure*}

Figure~\ref{fig:3} shows the DR cross sections obtained for the two dissociative states of the $^{2}\Pi$ symmetry corresponding to the (most important) $d$ partial wave of the incoming electron. We give here the {\it direct} - smooth cross section due to the direct capture of the electron into doubly excited dissociative states - and {\it total} or direct and indirect cross sections - infinite number of resonances, due to the temporary capture into highly excited Rydberg states. The $D_{1}$ dissociative state (black curve) has a more favourable crossing with the cation compared to $D_{2}$ (blue curve) leading thus to larger DR cross sections, although the vibronic coupling of $D_1$ is a factor of two smaller than the coupling of $D_2$. 

Figure~\ref{fig:4} shows the total DR cross sections associated to the $^{2}\Pi$ molecular states considering three scenarios for the partial waves of the incoming electron. Dark green stands for $p$ wave, blue for the $d$ and red for the $0.5\times p + 0.5\times d$ mixture. The overall behaviour of the three cross sections seems the same, at least starting from 100 meV collision energies. For very low collision energies we obtained smaller cross sections for the $p$ incident electrons.  

\begin{figure}\centering
		\includegraphics[width=0.95\columnwidth]{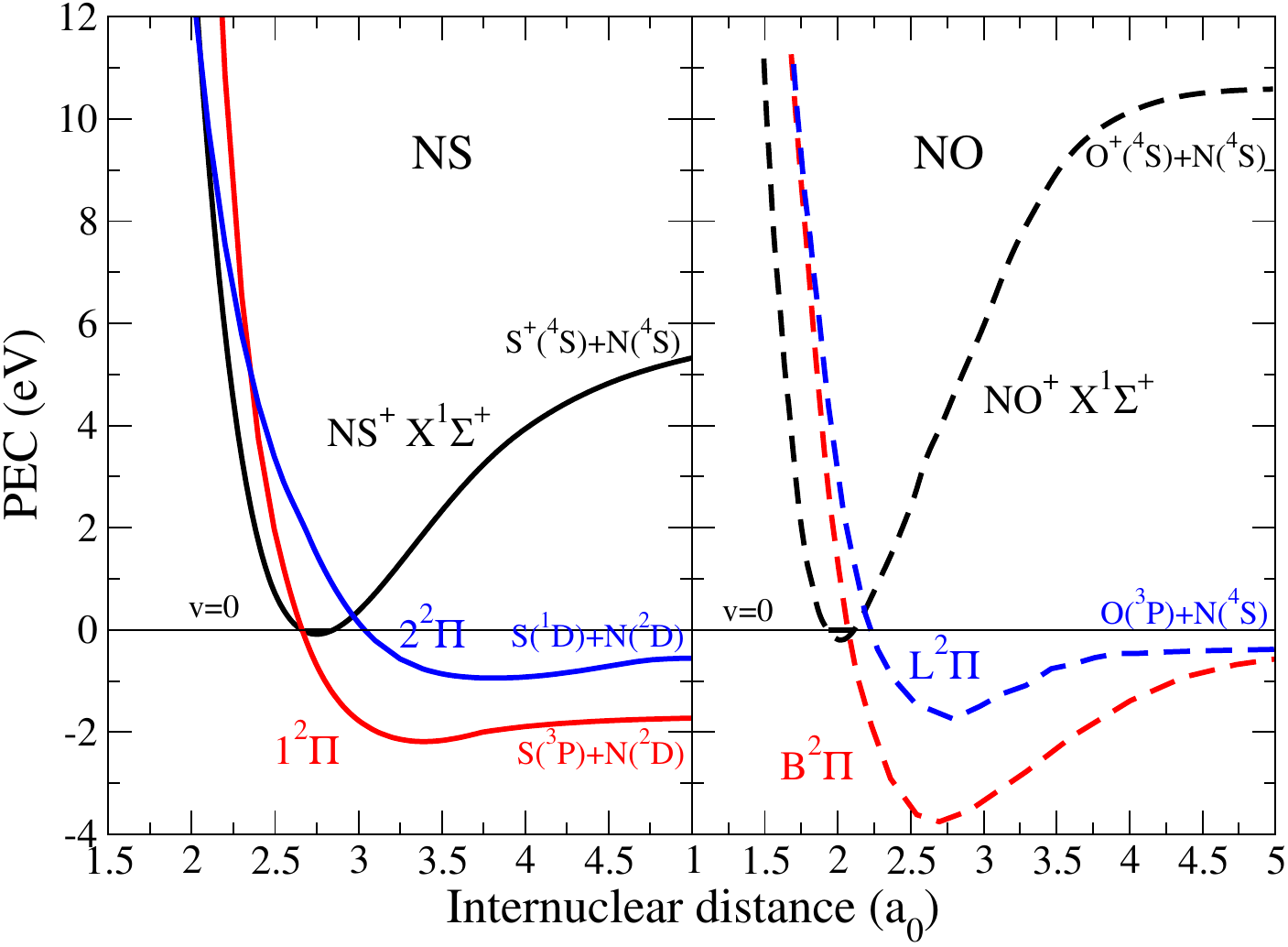}
    \caption{Potential energy curves of the NS$^+$/NS (left figure) and NO$^+$/NO molecular (right figure) systems relevant for low energy dissociative recombination compiled from~\cite{iacob2022} and ~\cite{Motapon2006} respectively. The solid curves stand for NS while the dashed ones for NO. The black curves show the ground electronic state of the target, while in red and blue we present the most important dissociative molecular states with $^2\Pi$ symmetry. 
    }
    \label{fig:6}
\end{figure}

The present nuclear dynamics study relies on the molecular data sets recently calculated in the framework of the R-matrix theory~\cite{iacob2022}. We found one ion core (NS$^+$ $X\,^1\Sigma^+$), three molecular symmetries (NS $^2\Sigma^{+}$, $^2\Pi$, and $^2\Delta$) and two or three partial waves for the incident electron for the three symmetries relevant for low-energy DR. Unfortunately, the electron structure and scattering calculations~\cite{iacob2022} provide global autoionization widths for the doubly excited dissociative resonant states, meaning that the electronic couplings are not resolved for the angular momentum of the incident electron. As a consequence, neither our cross sections nor rate coefficients are resolved. In order to overcome this we assume certain partial wave distributions for the vibronic couplings. This is the case for example in Fig.~\ref{fig:4}, where we have assumed either 100\% $p$ or $d$ waves or $0.5\times p + 0.5\times d$ mixture. 

Following this idea, we  made a detailed study considering $^2\Sigma^{+}$ and $^2\Pi$ symmetries of NS and $s$, $p$, $d$ or $p$ and $d$ partial waves for the two symmetries. The calculated rate coefficients are presented in Fig.~\ref{fig:5} for different partial wave distribution scenarios. The recommended rate coefficients for the DR of NS$^+$ $X^1\Sigma^+$ ($v_i^+=0,1$) are given in blue dashed-dotted lines labeled as M$(^2\Pi)+$M$(^2\Sigma^+)$. The grey area enclosed by the different scenarios can be regarded as uncertainties of MQDT calculations caused by the unresolved electronic couplings.

Due to the lack of any theoretical or experimental studies regarding the electron induced reactive processes of NS$^+$, we have instead compared our present NS$^+$ results with those previously obtained for NO$^+$~\cite{schneider2000,Motapon2006}, which due to its chemical similarity could be a proxy for NS$^+$. In Fig.~\ref{fig:6} we have compared the molecular states most relevant for DR of the NS$^+$/NS and NO$^+$/NO molecular systems. In black we represent the molecular state of the target cations, in red and blue the first two dissociative states of the neutral belonging to the $^2\Pi$ symmetry. The solid line stands for NS, while the dashed line for NO. 
The two target states are different, the NS$^+$ has a shallower but wider PEC, leading to less vibrational levels, thus fewer ionization channels. While the second pair of dissociative states (2 and L $^2\Pi$) presented in blue show similarities: they have unfavourable crossing points with the target states, similar shapes and asymptotic limits, we can see more important differences for the first two states (1 and B $^2\Pi$) given in red. These have different shapes and asymptotic limits, but most importantly the NO dissociative state has more favourable crossing with the cation. This latter is mostly responsible for the larger rate coefficients obtained for the DR of NO$^+$. 

\begin{figure}\centering
		\includegraphics[width=\columnwidth]{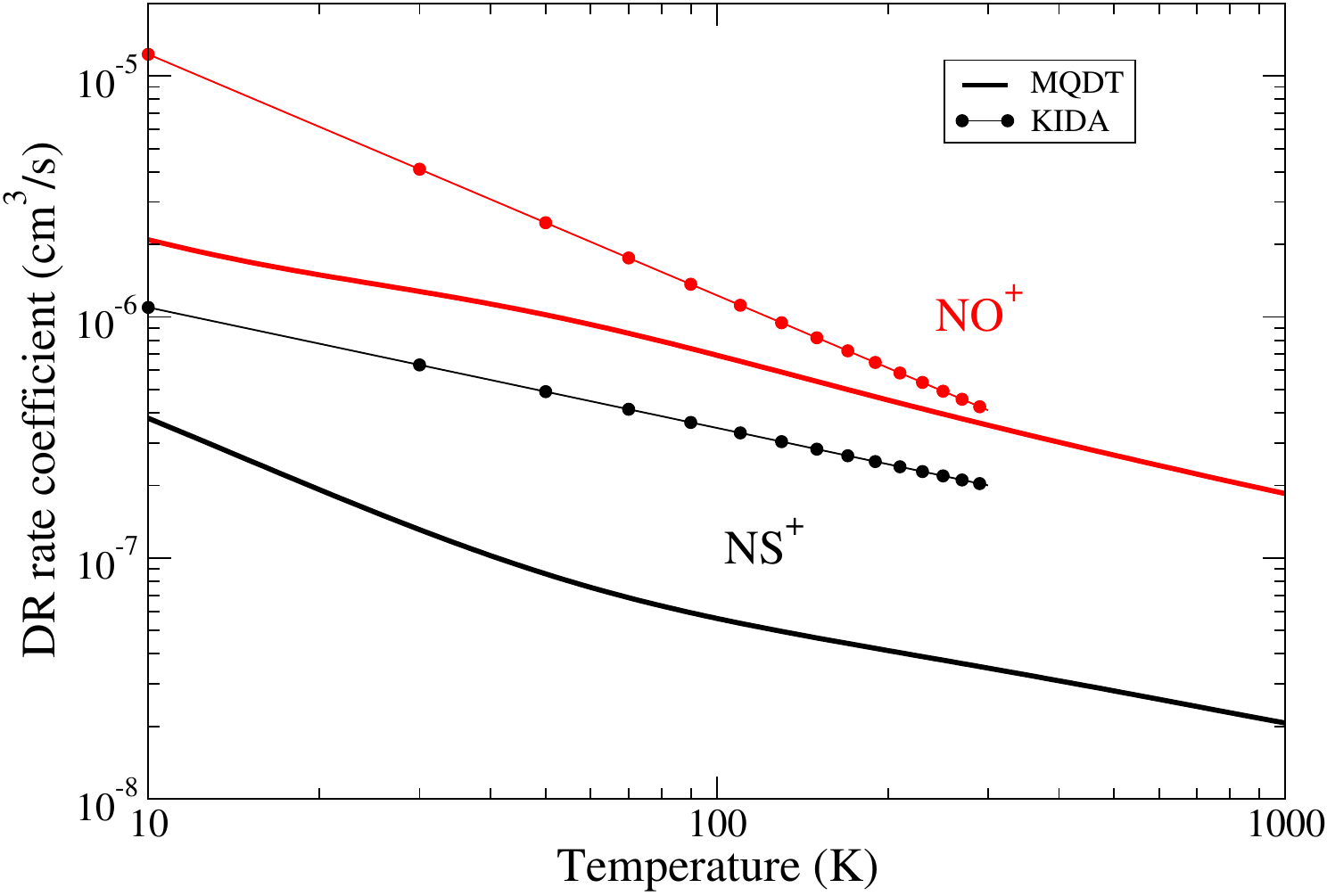}
    \caption{Dissociative recombination rate coefficients of NS$^+$ (in black) and NO$^+$~\cite{schneider2000,Motapon2006} (in red) molecular targets for $v_{i}^{+}=0$ compared with the rates available in the KIDA database~\cite{Wakelam2012}. 
    }
    \label{fig:7}
\end{figure}

The comparison between the NO$^+$ DR rate coefficient
and the NS$^+$ one for $v_{i}^{+}=0$
is performed in Fig.~\ref{fig:7}.
 We also compare here our MQDT-based rate coefficients with those displayed in the KIDA database \cite{Wakelam2012, prasad1980}.
 Our thermal rates are lower than the latter ones with a factor up to 7 for NO$^{+}$, and with a factor between 3 and 9 for NS$^{+}$.  
 It is stated that the KIDA 
 results are just estimates  \cite{Wakelam2012,prasad1980},
 without further information. As a conclusion, we consider that our rate coefficients obtained through MQDT for NS$^{+}$ are likely to be more reliable than those given by KIDA.
  As for NO$^{+}$, we note that our MQDT cross sections and rate coefficients \cite{schneider2000,Motapon2006} show excellent agreement with the experimental values 
 and should also be more reliable than those given by KIDA.

Finally, in order to simplify the use of our recombination rate coefficients for kinetic modelling, we have fitted their temperature dependence by using a simplified Arrhenius-type formula:
\begin{equation}
\alpha(T)=a_{0}\left(\frac{T}{300}\right)^{a_{1}}
\label{eq:Arrhenius}
\end{equation}
\noindent where $T$ is in Kelvin and $\alpha$ in cm$^{3}$s$^{-1}$. The fitting coefficients from the exponent became very small numbers, so they were omitted from the formula. The fitting parameters for the DR of the lowest 2 vibrational levels of the target, are summarized in table~\ref{tab:rate_DR1_fitt}.

\begin{table}
	\caption{Fitting parameters for the equation (\ref{eq:Arrhenius}), corresponding to the rate coefficients for DR on NS$^{+}$, with $v_{i}^{+}=0$ and $v_{i}^{+}=1$. Electron temperature range was divided in two regions and they vary between 10 and 1000 K.} 
	  \label{tab:rate_DR1_fitt}
	  \centering\footnotesize
	\begin{tabular}{ccccccc}
		\hline\hline
  \rule{0pt}{3ex} 
	Model & Temperature & $v_i^+$ & $a_{0}\times10^{7}$  & $a_{1}$ & RMS\\
          & (K) & & (cm$^{3}$s$^{-1})$ & \\
	    \hline
     \rule{0pt}{3ex}
      $\mbox{M}(^{2}\Pi)+\mbox{M}(^{2}\Sigma^{+})$ & 10$\le$ T $\le$ 33 & 0 & 0.1426 & -0.963555 & 0.0058\\
       & 33$<$ T $\le$ 78 &  & 0.23193 & -0.741244 &  0.0127\\
       & 78$<$ T $\le$ 1000 &  & 0.348726 & -0.43795 &  0.0075\\
\hline
     \rule{0pt}{3ex}
       $\mbox{M}(^{2}\Pi)+\mbox{M}(^{2}\Sigma^{+})$  & 10$\le$ T $\le$ 31 & 1 & 1.00126 & -0.509637 &  0.0089\\
       & 31 $<$ T $\le$ 160 &  & 1.22272 & -0.420919 & 0.0025\\
       & 160 $<$ T $\le$ 1000 &  & 1.32716 & -0.280971 & 0.0057\\
\hline\hline	
 \end{tabular}
\end{table}
\normalsize

According to the RMS values given in the table, the fitted parameters reproduce our MQDT rate coefficients within 1.3\% in the whole temperature range $10 < T< 1000$ K.  

\section{Conclusions}{\label{sec:conclusions}}

Using the step-wise multichannel quantum defect theory, we calculate cross sections and thermal rate coefficients for the dissociative recombination of electrons with NS$^{+}(X^{}\Sigma^{+})$ ions. The cross sections are computed within the energy range of $0.01$ meV to $1$ eV, and the corresponding thermal rate coefficients were evaluated for temperatures ranging from 10 to 1000 K, focusing on the two lowest vibrational levels of the ions.

Our model considers all pertinent electronic states and symmetries of the cation target, along with the relevant vibronic electronic couplings, by taking into account the quantum interference among the direct and indirect mechanisms.
We estimate the uncertainty of our MQDT calculations caused by the global electronic couplings unresolved in the angular momentum of the incident electron. Comparing our results with and those obtained for the similar molecular cation NO$^+$ and with values given in the KIDA database, suggests that the database values are too high. We suggest that rate coefficients obtained through MQDT for NS$^+$ and NO$^+$ are likely to be more reliable than those currently recommended in the KIDA database.
The relatively moderate dissociative recombination rate we evaluate explains - at least partially - the ubiquitous presence of NS$^+$ in interstellar space.

The numerical data, ready to be used in the kinetic modelling in astrochemistry and plasma physics is available upon request to the authors. 

\section*{Acknowledgements}
The authors acknowledge support provided by the F\'ed\'eration de Recherche Fusion par Confinement Magn\'etique (CNRS and CEA), La R\'egion Normandie, FEDER, and LabEx EMC3 through projects PTOLEMEE, Bioengine COMUE Normandie Universit\'e, the Institute for Energy, Propulsion and Environment (FR-IEPE), as well as the European Union through COST actions TUMIEE (CA17126), MW-Gaia (CA18104), and MD-GAS (CA18212). Additional support comes from l’Agence Universitaire de la Francophonie en Europe Centrale et Orientale (AUF ECO) through project CE/MB/045/2021 CiCaM -- ITER. The authors acknowledge the contribution of the Agence Nationale de la Recherche (ANR) through the MONA project. This work has received support from the Programme National "Physique et Chimie du Milieu Interstellaire" (PCMI) of CNRS/INSU with INC/INP co-funded by CEA and CNES. J.Zs.M. is grateful for financial support from the National Research, Development and Innovation Fund of Hungary, under the K18 and FK 19 funding schemes with project numbers K 128621 and FK 132989. JT and JZsM extend their appreciation to the Distinguished Guest Scientists Fellowship Programme-2022 of the Hungarian Academy of Sciences.
\section*{Data availability}
Upon a reasonable request, the data supporting this article will be provided by the corresponding author.

\bibliographystyle{unsrt}
\bibliography{NSp.bib}

\end{document}